\documentclass[a4paper,12pt]{article}
\usepackage[utf8]{inputenc}

\title{Fast Learning of Optimal Policy Trees}

\usepackage[sort,authoryear]{natbib}
\usepackage{amsmath}

\usepackage{pdflscape}
\usepackage{booktabs}
\usepackage{setspace}
\usepackage{longtable}
\usepackage[table]{xcolor} 
\usepackage{graphicx}
\usepackage{amssymb}
\usepackage{lscape}
\usepackage{longtable}
\usepackage{array}
\usepackage{makecell}
\usepackage{tikz}
\newcolumntype{P}[1]{>{\centering\arraybackslash}p{#1}}
\usepackage{caption}
\usepackage{subcaption}
\usepackage{threeparttable}
\usepackage[shortlabels]{enumitem}
\usepackage{float}
\usepackage[title]{appendix}
\usepackage{etoolbox}
\usepackage[multiple]{footmisc}
\usepackage{bbm}
\usepackage{authblk}
\usepackage{adjustbox}
\usepackage{graphicx}

\usepackage{lipsum}
\newcommand{\fpt}{\texttt{fastpolicytree}}
\newcommand{\pt}{\texttt{policytree}}

\author[1]{James Cussens}
\author[2]{Julia Hatamyar}
\author[3]{Vishalie Shah}
\author[4]{Noemi Kreif}

\affil[1]{University of Bristol}
\affil[2]{University of York}
\affil[3]{IQVIA}
\affil[4]{University of Washington}

\begin{document}
\maketitle
\doublespacing
\begin{abstract}

We develop and implement a version of the popular {\tt policytree} method (Athey and Wager, 2021) using discrete optimisation techniques. We test the performance of our algorithm in finite samples and find an improvement in the runtime of policytree learning by a factor of nearly 50 compared to the original version. We provide an {\tt R} package, {\tt fastpolicytree}, for public use. 
\end{abstract}

\section{Introduction}
The problem of learning optimal policy rules that map an individual covariate profile to a treatment decision has gained significant traction in recent years. Increasing availability of rich, large-scale data has contributed to this shift away from static policymaking towards a data-driven approach \citep{Amram2022-wt}. For example, in medicine, treatment decisions are increasingly personalised at the patient level according to patient characteristics and expected outcomes.

The existing literature on statistical methods for learning optimal policy rules is vast \citep{Manski2004-wf,Swaminathan2015-qm,Kitagawa2018-le,Van_der_Laan2015-ko,Luedtke2016-at,Luedtke2016-bi,Athey2021-uo}. Proposed approaches include those that assign a particular policy action (or treatment) to individuals whose expected benefit from being assigned to that treatment compared to a baseline treatment (estimated via the conditional average treatment effect (CATE) function) is positive, or those that directly select a policy rule that maximises expected outcomes across the population. Essentially, policy learning involves counterfactual estimation since we can only observe outcomes under the treatment actually received. 

In this paper, we consider the problem of learning optimal policy rules from an observational dataset of individuals $i=1,\ldots,n$, that are characterised by the form $(X_i,W_i,Y_i)$, where $X_i$ is a vector of covariates including confounders and effect modifiers, $W_i$ is a discrete treatment that could take one of $w=1,\ldots,m$ values, and $Y_i$ is the observed outcome. Our aim is to learn a policy rule $\pi$ that maps $X$ into a treatment decision $X \rightarrow \pi(x) \in \{1,\dots,m\}$ for policies in a pre-specified policy class $\pi \in \Pi$. The optimal policy rule $\pi^*$ is the rule that maximises (or minimises, depending on the objective function) the estimator of the policy value, often defined as the expected counterfactual mean reward $\mathbb{E}[\Gamma_{iw}(\pi(X_i))]$, where $\Gamma_{iw}$ is the estimated counterfactual reward for each observation $i$ under each policy action $w$. 

Depth-$k$ decision trees are a popular class for policy learning problems since they are able to generate interpretable and transparent outputs that describe the underlying model behind the policy decision. Decision trees recursively partition the covariate vector and assign a treatment decision to the partition that is aligned with a pre-specified tree depth, also known as the leaf node. The trained tree is then used to classify observations to a treatment decision by guiding them through the path of splits and into a leaf node according to their individual covariate profile. This tree-like structure means that the reasoning behind the decision rule is explainable, which is an often important consideration in personalised decision making. 

Traditionally, decision trees are trained using a top-down, greedy approach, meaning that the recursive algorithm starts at the root node and finds the locally optimal split at each partition, without considering the impact of future splits on overall tree performance. The problem with the greedy heuristic, however, is that the resulting tree may not capture all of the underlying characteristics of the data, leading to a suboptimal tree. A superior decision tree can be found by considering all future splits at each partition, thus achieving a globally optimal solution that maximises the objective function. Decision tree performance can be measured in terms of scalability (i.e. maintains performance as the dimensionality of data and tree depth increases), computational intensity and speed (i.e. generates a solution in a reasonable timeframe), and accuracy of optimality (i.e. correctly maximises the policy value estimator). 

In recent years, there have been some notable developments in decision tree-based methods for policy learning that are globally optimal. Some of these methods embed counterfactual estimation and policy learning within the same decision tree, while others separate these two tasks by first estimating the counterfactuals using an appropriate data model, and then training a decision tree on these estimates to learn the optimal policy rule \citep{Amram2022-wt}. We focus on methods that use the latter approach since we are only interested in the policy optimisation step in this work. \citet{Zhou_2023} develop an implementation of an exhaustive tree-search based algorithm that finds an exact optimal tree. However, their tree struggles to scale well beyond shallow trees and small datasets. Based on the work of \citet{Bertsimas2017-bs}, they also consider formulating the policy learning problem as a mixed integer programme (MIP), which can be solved using commercial solvers (e.g.\ Gurobi \citep{gurobi}), however they encounter similar scaling issues using this approach. \citet{Amram2022-wt} develop an alternative implementation of the exact tree search that uses coordinate descent to train the decision trees by repeatedly optimising each tree split until no further improvement in the overall objective function can be achieved.

\section{Our implementation}

Our implementation is available as an R package called \fpt{}. Its goal is simply to learn policy trees more quickly than the existing \pt{} R package. We start with an account of the tree-building approach found in \pt.

\subsection{\pt}
\label{sec:pt}

The \texttt{tree\_search} function provided by \pt{}, finds an optimal policy tree (of a given maximal depth) for any given subset of the covariate and reward data. It does this via a recursive algorithm. Given any subset of the data and depth limit $d$ as input, 
\texttt{tree\_search} considers every possible split (each split determined by a choice of covariate and covariate splitting value) thus creating a `left' subset of the data and a `right' subset. Once an optimal policy tree for both the left and right subsets have been found (via recursion with depth limit $d-1$), we then know the best possible reward for the given split. Since all splits are considered, this allows the algorithm to determine the optimal policy tree for its input. 

A key issue is how these subsets are implemented. In \pt{} each subset is stored as $p$ sorted sets where $p$ is the number of covariates. The sorted set for covariate $j$ has (the indices) of the datapoints in the subset sorted according to their values on covariate $j$. A major advantage of this approach is that when generating successive splits for covariate $j$ one can quickly find the datapoints to move from the `right' to the `left'. A disadvantage is that each such move requires updating $p$ sorted sets.

\citet{Sverdrup2020-yj} state that each sorted set is implemented as a binary search tree which would allow datapoints to be added and removed in $O(\log n)$, where $n$ is the size of the set. However, \pt{} actually implements sorted sets as a sorted vector. The Boost container data type \texttt{flat\_set} \citep{flatset} is used which does allow faster iteration through the set than using a tree but, addition and removal are slower. To find where a new element should be inserted in a flat set takes $O(\log n)$ time, but then it is also necessary to shift elements to make room for the new element---which takes $O(n)$ time. Examining \pt{} using the Linux perf tool we find that it spends roughly 25\%--30\% of its time shifting elements. 
Note that \pt{} has an important optimisation when looking for depth $d=1$ trees. In this case, there is no need to maintain $p$ sorted sets (since there will be no further splits) and so a faster method is used. 

\subsection{Discrete optimisation}
\label{sec:do}

Before describing how \fpt{} is implemented it will be useful to explain some general principles of \emph{discrete optimisation}. A discrete optimisation task is to find the optimal member of some finite set, so exact policy tree learning is such a task. Typically the set is very large so simply inspecting each member of the set (\emph{brute-force search}) is impracticable. Instead methods which exploit the structure of the problem are used.
The recursive approach taken by \pt{} does exploit the tree structure of policy trees by breaking the problem down into smaller subproblems which can be solved independently: finding the trees which are optimal for the subsets on the two side of a split can be done independently.

A key technique in discrete optimisation is the use of \emph{bounds}. Suppose that at some point in the search for an optimal policy tree we are about to find the optimal left and right policy trees for some split. Suppose also that we have a record of the best policy tree found so far (known as the \emph{incumbent}). If we have an upper bound on the objective value (i.e.\ the reward) of the optimal policy tree for the current split and this upper bound is \emph{below} the objective value of the incumbent, then there is no point finding the optimal tree for the current split, since we know it will not beat the incumbent. The key to the effective use of bounds is to be able to compute them reasonably quickly and for the bounds to be reasonably `tight'---not too far from the (unknown) value being bounded.

\subsection{\fpt}
\label{sec:fpt}

To explain \fpt{} it is useful to formalise the policy tree learning problem and introduce some notation. We are considering the problem of finding an optimal policy tree
for a set of \emph{units}, where a unit is a collection of covariate
values together with a reward for each possible action. Let $N$ be the set $\{1,\dots,n\}$ of
units. Let $A$ be the set of available actions.  Let ${\cal F}_d$
represent the set of functions $f:N \rightarrow A$ representable by a
policy tree of depth $d$. $f(i)$ assigns an action to unit $i$, so it
is an \emph{action assignment function}. Let $S$ be the set of splits
available, each split $s \in S$ being defined by a choice of covariate
$j$ and a value $v$ for that covariate, so each $s \in S$ is
determined by a pair $s=(j,v)$. Let $x_{ij}$ be the value of
covariate $j$ for unit $i$. If a unit $i$ is such that
$x_{ij} \leq v$ then we say the unit is sent left by the split $s=(j,v)$,
otherwise it is sent right.

Let $s_{L}(N)$ (resp.\ $s_{R}(N)$) be the units sent left (resp.\
right) when unit-set $N$ is split using split $s \in S$. Let $r(i,a)$
be the reward for unit $i$ when it is assigned action $a$. For any
action assignment function $f$ define
$R(f,N) := \sum_{i \in N} r(i,f(i))$, so $R(f,N)$ is the reward when
using $f$ to assign actions to all individuals in $N$. Define
$f^{*}_{d,N} := \arg\max_{f \in {\cal F}_d} R(f,N)$, so $f^{*}_{d,N}$
is the best function in ${\cal F}_d$ (the best depth $d$ policy tree)
to use to assign actions to all individuals in $N$. Given $d$ and $N$
our goal is to find $f^{*}_{d,N}$.

Abbreviate $R(f^{*}_{d,N},N)$ to $R^{*}_{d,N}$, so $R^{*}_{d,N}$ is the best
possible reward---the reward resulting from using the best tree
$f^{*}_{d,N}$---for a given $d$ and $N$. We have the basic recursion 
for $d>0$
\begin{equation}
  \label{eq:basicrecursion}
  R^{*}_{d,N}
  = \max_{s \in S}   R^{*}_{d-1,s_{L}(N)} + R^{*}_{d-1,s_{R}(N)} 
\end{equation}
The base case is:
\begin{equation}
  \label{eq:base}
    R^{*}_{0,N} =  \max_{a \in A} \sum_{i \in N}  r(i,a)
\end{equation}
(\ref{eq:basicrecursion}) and (\ref{eq:base}) lead to a simple
algorithm for finding $f^{*}_{d,N}$: compute $R^{*}_{d-1,s_{L}(N)} +
R^{*}_{d-1,s_{R}(N)}$ by recursion for each split $s \in S$ and record
which has the highest reward. By recording the maximising
split during the recursive computation we can recover the policy tree
which produces this maximal reward. This is the algorithm used by \pt.

\subsubsection{Using bounds}
\label{sec:usingbounds}

Suppose we have found the optimal policy tree $f^{*}_{d,N_{1}}$ for some set of units $N_{1}$ and depth $d$. Abbreviate $R(f^{*}_{d,N_{1}})$, the reward for this optimal tree, to $R^{*}_{d,N_{1}}$. Now suppose we add some new units $N_3$ to $N_1$ such as would happen to a `left' set if we increased the splitting value for some covariate. We claim that the following upper bound (\ref{eq:simpleleft}) on 
$R^{*}_{d,N_{1}\dot\cup N_{3}}$ is valid. ($\dot\cup$ represents disjoint union.)
\begin{equation}
  \label{eq:simpleleft}
  R^{*}_{d,N_{1}\dot\cup N_{3}}  \leq R^{*}_{d,N_{1}} +    \sum_{i \in
  N_{3}} \max_{a \in A} r(i,a)
\end{equation}
The upper bound (\ref{eq:simpleleft}) makes the following (intuitively reasonable) claim: for units $N_{1}\dot\cup N_{3}$ and tree depth $d$, the best reward we can hope for is the best reward possible for a depth $d$ tree for units $N_1$ alone, plus the best possible reward for each unit in $N_3$. We can prove (\ref{eq:simpleleft}) by assuming it false and deriving a contradiction.
\begin{eqnarray*}
  R^{*}_{d,N_{1}\dot\cup N_{3}} & > &R^{*}_{d,N_{1}} +    \sum_{i \in
    N_{3}} \max_{a \in A} r(i,a)\\
\Leftrightarrow  R(f^{*}_{d,N_{1}\dot\cup N_{3}},N_{1}) + R(f^{*}_{d,N_{1}\dot\cup N_{3}},N_{3})  & > &R^{*}_{d,N_{1}} +    \sum_{i \in
    N_{3}} \max_{a \in A} r(i,a)\\
\Leftrightarrow  R(f^{*}_{d,N_{1}\dot\cup N_{3}},N_{1}) & > & R^{*}_{d,N_{1}} +    \sum_{i \in
    N_{3}} \left[ \max_{a \in A} r(i,a) - R(f^{*}_{d,N_{1}\dot\cup N_{3}},i)\right]
\end{eqnarray*}
The last line is a contradiction since each term in $\sum_{i \in N_{3}} [ \max_{a \in A} r(i,a) - R(f^{*}_{d,N_{1}\dot\cup N_{3}},i)]$ is non-negative and so the inequality asserts that there is a tree (namely $f^{*}_{d,N_{1}\dot\cup N_{3}}$) which has a strictly higher reward for $N_1$ than the optimal tree for $N_1$.

Now consider removing some units $N_3$ from a set of units $N_2$ as would happen to a `right' set if we increased the splitting value for some covariate. We claim that the following upper bound (\ref{eq:simpleright}) on 
$R^{*}_{d,N_{2}\setminus N_{3}}$ is valid.
\begin{equation}
  \label{eq:simpleright}
    R^{*}_{d,N_{2}\setminus N_{3}} \leq R^{*}_{d,N_{2}} - \sum_{i \in
    N_{3}} \min_{a \in A} r(i,a)
\end{equation}
The upper bound (\ref{eq:simpleright}) makes the following claim:  for units $N_{2}\setminus N_{3}$ (i.e.\ the set of units in $N_2$ but not $N_3$) and tree depth $d$, the best reward we can hope for is the best reward possible for a depth $d$ tree for units $N_2$, minus the lowest possible reward for each unit in $N_3$. We can prove (\ref{eq:simpleright}) by assuming it false and deriving a contradiction.
\begin{eqnarray*}
    R^{*}_{d,N_{2}\setminus N_{3}} & > & R^{*}_{d,N_{2}} - \sum_{i \in
    N_{3}} \min_{a \in A} r(i,a) \\
    \Leftrightarrow  R(f^{*}_{d,N_{2}\setminus N_{3}},N_{2}) - R(f^{*}_{d,N_{2}\setminus N_{3}},N_{3}) & > & R^{*}_{d,N_{2}} - \sum_{i \in
    N_{3}} \min_{a \in A} r(i,a) \\
\Leftrightarrow  R(f^{*}_{d,N_{2}\setminus N_{3}},N_{2}) & > & R^{*}_{d,N_{2}} + \sum_{i \in
    N_{3}} \left[  R(f^{*}_{d,N_{2}\setminus N_{3}},i) - \min_{a \in A} r(i,a)\right]
\end{eqnarray*}
The last line is a contradiction since each term in $\sum_{i \in
    N_{3}} [  R(f^{*}_{d,N_{2}\setminus N_{3}},i) -  \min_{a \in A} r(i,a)]$ is non-negative and so the inequality asserts that there is a tree (namely $f^{*}_{d,N_{2}\setminus N_{3}}$) which has a strictly higher reward for $N_2$ than the optimal tree for $N_2$.

\fpt{} uses the bounds (\ref{eq:simpleleft}) and (\ref{eq:simpleright}) as follows. Suppose we are trying to find $f^{*}_{d,N}$, the optimal policy tree of depth $d$ for a set of units $N$. Suppose that we have already found $f^{*}_{d-1,N_{1}}$, and $f^{*}_{d-1,N_{2}}$, where $N = N_{1} \dot\cup N_{2}$ so we have the optimal reward when $N$ is split into $N_1$ and $N_2$---namely $R^{*}_{d-1,N_{1}} + R^{*}_{d-1,N_{2}}$. Suppose that later on we are considering a different split of $N$ where $N_3$ has been removed from $N_2$ and added to $N_{1}$ (such as is obtained by increasing the splitting value for some covariate). The optimal reward for this split is $R^{*}_{d-1,N_{1}\dot\cup N_{3}} +
R^{*}_{d-1,N_{2}\setminus N_{3}}$. From (\ref{eq:simpleleft}) and (\ref{eq:simpleright}) we have:
\begin{equation}
\label{eq:usingbound}
  R^{*}_{d-1,N_{1}\dot\cup N_{3}} +
R^{*}_{d-1,N_{2}\setminus N_{3}} \leq R^{*}_{d-1,N_{1}} + R^{*}_{d-1,N_{2}} + \sum_{i \in
  N_{3}} \max_{a \in A} r(i,a)   - \min_{a \in A} r(i,a) 
\end{equation}
So given that we already have $R^{*}_{d-1,N_{1}} + R^{*}_{d-1,N_{2}}$ the RHS of (\ref{eq:usingbound}) provides a cheaply computable upper bound on  $R^{*}_{d-1,N_{1}\dot\cup N_{3}} + R^{*}_{d-1,N_{2}\setminus N_{3}}$. \fpt{} computes and stores $\max_{a \in A} r(i,a)$ and $\min_{a \in A} r(i,a)$ for every unit $i$ before starting the search for an optimal policy tree, so $\sum_{i \in
  N_{3}} \max_{a \in A} r(i,a)   - \min_{a \in A} r(i,a)$ is quick to compute. If the current incumbent for set $N$ and depth $d$ has a reward above this upper bound then \fpt{} does not waste time computing $R^{*}_{d-1,N_{1}\dot\cup N_{3}} +
R^{*}_{d-1,N_{2}\setminus N_{3}}$ since we know it cannot be an optimal reward.

\subsubsection{Alternative set implementation}
\label{sec:altset}

\fpt{} implements sets of units in two ways. Method~1 is the sorted vector (one for each covariate) approach taken by \pt, although this is implemented as a C struct rather than using the Boost C++ library. Method~2 uses a single set rather than a set for each covariate. Just before iterating over splits for covariate $j$ this set is sorted according to the values of covariate $j$. An advantage of this approach is that moving a unit from right to left (as we increase the splitting value of covariate $j$) is quick and does not require another $p-1$ sorted sets to be updated. The disadvantage, of course, is the sorting mentioned above. 
If the number of distinct values of a covariate is below a certain threshold (currently 30) then \emph{counting sort} is used for sorting, otherwise \emph{radix sort} is used. See \citet{cormen90:_introd_algor} for a description of these two sorting algorithms.

If, when using Method~2, a covariate is found to have only 2 distinct values in the data (such as would be the case for a binary variable) then each unit is directly assigned to the left or right set as appropriate without sorting the set in the normal way.

Method~1 and Method~2 have competing benefits. At present, if most covariates have no more than 30 distinct values then Method~2 is used, otherwise Method~1 is used.

\subsubsection{Caching}

\begin{figure}
\centering
\begin{tikzpicture}
    \node (root) at (2,2) {$x_{1}\leq 2.4$};
    \node (left) at (1,1) {$x_{2} \leq 3.7$};
    \node (right) at (3,1) {\dots};
    \node (leftleft) at (0,0) {X};
    \node (leftright) at (2,0) {\dots};
    \draw (root) -- (left);
    \draw (root) -- (right);
    \draw (left) -- (leftleft);
    \draw (left) -- (leftright);
    \node (root) at (7,2) {$x_{2}\leq 3.7$};
    \node (left) at (6,1) {$x_{1} \leq 2.4$};
    \node (right) at (8,1) {\dots};
    \node (leftleft) at (5,0) {X};
    \node (leftright) at (7,0) {\dots};
    \draw (root) -- (left);
    \draw (root) -- (right);
    \draw (left) -- (leftleft);
    \draw (left) -- (leftright);
\end{tikzpicture}
\caption{Two partially constructed policy trees which have the same tree to find at position X. Units satisfying an inequality at a tree node are sent to its left branch, and those not satisfying it are sent right.}
\label{fig:cacheex}
\end{figure}
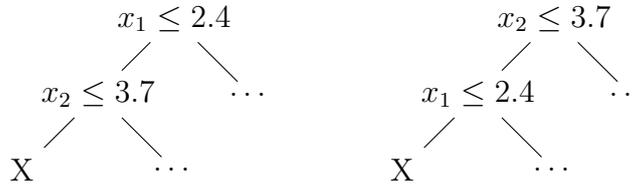

Consider the two partially constructed policy trees displayed in Fig~\ref{fig:cacheex}. The left-hand tree illustrates the situation where we are looking for an optimal tree with top-level split $x_{1} \leq 2.4$ and are looking for an optimal tree with split $x_{2} \leq 3.7$ for those units satisfying $x_{1} \leq 2.4$. The right-hand tree is the same except with $x_{2} \leq 3.7$ at the top and $x_{1} \leq 2.4$ below. It is clear that the set of units $N$ reaching node X is the same in both cases: those where both $x_{1} \leq 2.4$ and $x_{2} \leq 3.7$. Since X is at the same depth $d$ in both cases it follows that the optimal policy tree is the same ($f^{*}_{d,N}$) in both cases.

For each set of units $N$ and depth $d$, once \fpt{} has found the optimal policy tree $f^{*}_{d,N}$, it is stored in a cache (unless the cache has got too big, at present the cache is limited to have at most 1,000,000 trees). If at a later point the algorithm is required to find $f^{*}_{d,N}$ again, it is simply retrieved from the cache.

\subsubsection{Miscellaneous optimisations}

The previous three sections give the main optimisations found in \fpt: using bounds, allowing for a different implementation of sets and caching. There are also some other optimisations:
\begin{description}
    \item[Perfect trees] If, while attempting to find the optimal policy tree $f^{*}_{d,N}$ for some set of units $N$ and depth $d$, we find a tree that assigns $\max_{a \in A} r(i,a)$ (i.e.\ the best action) to each unit $i \in N$, then we know we have an optimal tree (since no tree, of any depth, can have higher reward) and we can stop searching further. We call such trees \emph{perfect}.
    \item[Pre-allocating memory] During the course of its search for an optimal tree, \fpt{} constructs many set of units $N$, many (mostly temporary) policy trees and some other temporary arrays of values. Allocating memory for such data structures can be expensive, so \fpt{} allocates the necessary memory once at the start of the search and frees it at the end. So rather than allocating memory to store a new tree, pre-allocated space is used and the new tree overwrites any tree previously stored in that space.
\end{description}

\section{Simulation study}

In this section, we perform several experiments on synthetic data to test the relative performance of our implementation of the optimal decision tree for policy learning against existing implementations described in the previous section. Our simulation setup has been inspired by \citet{Zhou_2023}.

In our experimental set up, we generate variations of the following parameters: the sample size $n$ (500, 1000 and 2000), the dimension $p$ of the covariate vector $X$ (5 and 10), the type of covariate vector (continuous or discrete), the number of available treatments $m$ (2 and 3 ), and the decision tree depth $d$ (2 and 3). We simulate the covariate vector $X$ by drawing variables $X_1,\ldots,X_p$ from the standard normal distribution for continuous covariate data, and from the Bernoulli(0.5) distribution for discrete covariate data. The treatment variable $W$ is a categorical variable that can take $m$ values, and is generated by randomly sampling a vector of two or more elements depending on the value of $m$, corresponding to a setting of fully randomised treatment without confounding. The outcome variable is a function of $X$, $W$ and an error term $\epsilon$ drawn from a uniform distribution, as follows:
\begin{equation}
    Y = X_1 + X_{2}I(W=1) + X_{3}I(W=m) + \epsilon.
\end{equation}
This data generating process illustrates a setting with treatment effect heterogeneity, the effect of the treatment varying with the baseline covariate $X_3$. As $w$ changes (i.e. treatment shifts away from baseline $W=1$), the treatment effect is proportional to $X_3$ and scales with $w-1$. For each variation of the data generating process (DGP), we perform 10 repetitions. 

Our performance measures include the runtime for training the decision tree (denoted $T$, measured using elapsed CPU time) and the estimator of the policy value under the learned policy rule $\mathbb{E}[\Gamma_{iw}(\pi(X_i))]$, which requires estimates of the reward matrix $\Gamma_{iw}$. For ease of implementation, we estimate $\Gamma_{iw}$ by following the procedure used by \citet{Athey2021-uo} and \citet{Zhou_2023} from the causal inference and policy evaluation literature, which generates estimates of heterogeneous treatment effects (via the CATE function) that can be used (along with estimates of the outcome and treatment models) to construct $\Gamma_{iw}$, as follows:
\begin{equation}
    \Gamma_{iw} = \hat{\mu}_{w}(X_i) + \frac{Y_i-\hat{\mu}_{w}(X_i)}{\hat{e}_{w}(X_i)}\cdot \mathbbm{1}\{W_i=w\},
\end{equation}
where $\hat{\mu}_{w}(X_i)$ is the estimated counterfactual response surface for each $i$ under each treatment $w$, and $\hat{e}_{w}(X_i)$ is the estimated probability that $i$ is assigned to treatment $w$. These parameters are estimated by fitting a causal forest, which is a causal adaption of the random forest prediction algorithm \citep{Athey2019-qf}. 

We compare our implementation of policy trees to those used in \citet{Zhou_2023} and \citet{Athey2021-uo} (i.e. \texttt{policytree}\footnote{https://github.com/grf-labs/policytree} developed by \citet{Sverdrup2020-yj}).


Our simulation procedure for each DGP can be described in the following steps:
\begin{enumerate}
    \item For each repetition:
    \begin{enumerate}
        \item Generate data according to sample size ($n$), dimension of the covariate vector ($p$), data type, number of treatments ($m$) and decision tree depth ($d$).
        \item Fit a causal forest (when $m=2$) or a multi-arm causal forest (when $m>2$).
        \item Compute the rewards matrix $\Gamma_{iw}$.
        \item For each implementation of the policy tree:
        \begin{enumerate}
            \item Train the tree (with $\Gamma_{iw}$ and $X_i$ as inputs) and store the runtime $T$.
            \item Generate predictions of the policy rule $\pi$ for each $i$.
            \item Estimate and store the policy value $\mathbb{E}[\Gamma_{iw}(\pi(X_i))]$.
        \end{enumerate}
    \end{enumerate}
    \item Take averages of the stored values across repetitions.
\end{enumerate}

\clearpage
\section{Results}

\subsection{Comparison to Policytree with default parameters}

In this section we compare \pt{} to \fpt{} where \pt{} has all its parameters set to default values, except tree depth (where appropriate). When \pt{} is run in this way both it and \fpt{} find an optimal policy tree for the given depth. We have checked that the rewards of the trees found by \pt{} and \fpt{} are identical in all our simulations. It follows that the time taken to produce an optimal policy tree is the only quantity of interest. Our timing results are presented in Table~\ref{tab:ptdefault}

\singlespacing

\setlength{\tabcolsep}{4pt} 

\begin{longtable}[t]{ccccccccccc}
\caption[]{Simulation results: \pt{} with default parameters ({\bf PT}) vs \fpt{} ({\bf FPT}). SD is the standard deviation of the time over 100 simulation repetitions. Data: covariate data ($X$).}\\
\toprule
\multicolumn{5}{c}{\textbf{ }} & \multicolumn{2}{c}{\textbf{PT}} & \multicolumn{2}{c}{\textbf{FPT}}\\
\cmidrule{6-7} \cmidrule{8-9}
N & p & Acts & Depth & Data  & Time & SD & Time & SD\\
\midrule
\endfirsthead
\caption[]{Simulation results \textit{(continued)}}\\
\toprule
\multicolumn{5}{c}{\textbf{ }} & \multicolumn{2}{c}{\textbf{PT}} & \multicolumn{2}{c}{\textbf{FPT}}\\
\cmidrule{6-7} \cmidrule{8-9}
N & p & Acts & Depth & Data  & Time & SD & Time & SD\\
\midrule
\endhead

\endfoot
\bottomrule
\multicolumn{9}{l}{\rule{0pt}{1em}\textit{Note: Results are averaged across 100 simulation repetitions.}}\\
\endlastfoot
1000 & 10 & 2 & 2 & discrete & 0.026 sec & 0.001 sec & 0.002 sec & 0 sec \\ 
  5000 & 10 & 2 & 2 & discrete & 0.445 sec & 0.017 sec & 0.006 sec & 0.001 sec \\ 
  10000 & 10 & 2 & 2 & discrete & 1.727 sec & 0.052 sec & 0.011 sec & 0.001 sec \\ 
    \addlinespace
  1000 & 30 & 2 & 2 & discrete & 0.257 sec & 0.009 sec & 0.008 sec & 0.001 sec \\ 
  5000 & 30 & 2 & 2 & discrete & 4.544 sec & 0.133 sec & 0.031 sec & 0.001 sec \\ 
  10000 & 30 & 2 & 2 & discrete & 16.621 sec & 0.797 sec & 0.059 sec & 0.003 sec \\ 
    \addlinespace
  1000 & 60 & 2 & 2 & discrete & 1.252 sec & 0.043 sec & 0.025 sec & 0.001 sec \\ 
  5000 & 60 & 2 & 2 & discrete & 19.75 sec & 1.287 sec & 0.108 sec & 0.005 sec \\ 
  10000 & 60 & 2 & 2 & discrete & 2.382 min & 1.122 min & 0.212 sec & 0.009 sec \\ 
    \addlinespace
  1000 & 10 & 3 & 2 & discrete & 0.028 sec & 0.001 sec & 0.004 sec & 0 sec \\ 
  5000 & 10 & 3 & 2 & discrete & 0.444 sec & 0.014 sec & 0.007 sec & 0.001 sec \\ 
  10000 & 10 & 3 & 2 & discrete & 1.728 sec & 0.056 sec & 0.013 sec & 0.001 sec \\ 
    \addlinespace
  1000 & 30 & 3 & 2 & discrete & 0.259 sec & 0.01 sec & 0.011 sec & 0.001 sec \\ 
  5000 & 30 & 3 & 2 & discrete & 4.54 sec & 0.153 sec & 0.036 sec & 0.002 sec \\ 
  10000 & 30 & 3 & 2 & discrete & 15.798 sec & 0.576 sec & 0.068 sec & 0.003 sec \\ 
    \addlinespace
  1000 & 60 & 3 & 2 & discrete & 1.243 sec & 0.044 sec & 0.026 sec & 0.001 sec \\ 
  5000 & 60 & 3 & 2 & discrete & 18.59 sec & 0.685 sec & 0.122 sec & 0.005 sec \\ 
  10000 & 60 & 3 & 2 & discrete & 1.958 min & 46.829 sec & 0.246 sec & 0.011 sec \\ 
    \addlinespace
  1000 & 10 & 10 & 2 & discrete & 0.03 sec & 0.002 sec & 0.006 sec & 0.001 sec \\ 
  5000 & 10 & 10 & 2 & discrete & 0.453 sec & 0.019 sec & 0.011 sec & 0.007 sec \\ 
  10000 & 10 & 10 & 2 & discrete & 1.724 sec & 0.079 sec & 0.018 sec & 0.002 sec \\ 
    \addlinespace
  1000 & 30 & 10 & 2 & discrete & 0.263 sec & 0.009 sec & 0.011 sec & 0.001 sec \\ 
  5000 & 30 & 10 & 2 & discrete & 4.542 sec & 0.217 sec & 0.051 sec & 0.003 sec \\ 
  10000 & 30 & 10 & 2 & discrete & 15.924 sec & 0.679 sec & 0.109 sec & 0.006 sec \\ 
    \addlinespace
  1000 & 60 & 10 & 2 & discrete & 1.267 sec & 0.059 sec & 0.034 sec & 0.002 sec \\ 
  5000 & 60 & 10 & 2 & discrete & 18.65 sec & 0.914 sec & 0.188 sec & 0.01 sec \\ 
  10000 & 60 & 10 & 2 & discrete & 1.519 min & 29.649 sec & 0.414 sec & 0.022 sec \\ 
    \addlinespace
  1000 & 10 & 2 & 3 & discrete & 0.196 sec & 0.009 sec & 0.005 sec & 0.001 sec \\ 
  5000 & 10 & 2 & 3 & discrete & 2.677 sec & 0.123 sec & 0.012 sec & 0.001 sec \\ 
  10000 & 10 & 2 & 3 & discrete & 10.347 sec & 0.482 sec & 0.023 sec & 0.001 sec \\ 
    \addlinespace
  1000 & 30 & 2 & 3 & discrete & 5.316 sec & 0.234 sec & 0.068 sec & 0.003 sec \\ 
  5000 & 30 & 2 & 3 & discrete & 1.399 min & 3.891 sec & 0.317 sec & 0.013 sec \\ 
  10000 & 30 & 2 & 3 & discrete & 4.764 min & 11.642 sec & 0.627 sec & 0.025 sec \\ 
    \addlinespace
  1000 & 60 & 2 & 3 & discrete & 49.23 sec & 1.844 sec & 0.546 sec & 0.021 sec \\ 
  5000 & 60 & 2 & 3 & discrete & 11.628 min & 36.105 sec & 2.722 sec & 0.11 sec \\ 
  10000 & 60 & 2 & 3 & discrete & 40.86 min & 3.869 min & 5.452 sec & 0.222 sec \\ 
    \addlinespace
  1000 & 10 & 3 & 3 & discrete & 0.196 sec & 0.008 sec & 0.007 sec & 0.001 sec \\ 
  5000 & 10 & 3 & 3 & discrete & 2.627 sec & 0.113 sec & 0.016 sec & 0.002 sec \\ 
  10000 & 10 & 3 & 3 & discrete & 10.019 sec & 0.473 sec & 0.027 sec & 0.001 sec \\ 
    \addlinespace
  1000 & 30 & 3 & 3 & discrete & 5.234 sec & 0.231 sec & 0.079 sec & 0.003 sec \\ 
  5000 & 30 & 3 & 3 & discrete & 1.37 min & 3.783 sec & 0.381 sec & 0.015 sec \\ 
  10000 & 30 & 3 & 3 & discrete & 4.607 min & 14.405 sec & 0.756 sec & 0.03 sec \\ 
    \addlinespace
  1000 & 60 & 3 & 3 & discrete & 49.712 sec & 2.045 sec & 0.635 sec & 0.023 sec \\ 
  5000 & 60 & 3 & 3 & discrete & 11.347 min & 39.692 sec & 3.223 sec & 0.128 sec \\ 
  10000 & 60 & 3 & 3 & discrete & 41.27 min & 3.539 min & 6.49 sec & 0.251 sec \\ 
    \addlinespace
  1000 & 10 & 10 & 3 & discrete & 0.201 sec & 0.009 sec & 0.007 sec & 0.001 sec \\ 
  5000 & 10 & 10 & 3 & discrete & 2.65 sec & 0.128 sec & 0.023 sec & 0.002 sec \\ 
  10000 & 10 & 10 & 3 & discrete & 10.048 sec & 0.5 sec & 0.052 sec & 0.004 sec \\ 
    \addlinespace
  1000 & 30 & 10 & 3 & discrete & 5.401 sec & 0.245 sec & 0.111 sec & 0.005 sec \\ 
  5000 & 30 & 10 & 3 & discrete & 1.38 min & 3.767 sec & 0.619 sec & 0.028 sec \\ 
  10000 & 30 & 10 & 3 & discrete & 4.598 min & 14.105 sec & 1.343 sec & 0.068 sec \\ 
    \addlinespace
  1000 & 60 & 10 & 3 & discrete & 50.724 sec & 2.158 sec & 0.91 sec & 0.037 sec \\ 
  5000 & 60 & 10 & 3 & discrete & 11.399 min & 40.259 sec & 5.182 sec & 0.234 sec \\ 
  10000 & 60 & 10 & 3 & discrete & 39.247 min & 3.137 min & 11.397 sec & 0.564 sec \\ 
  \addlinespace
  \hline
  \multicolumn{6}{l}{\textbf{Panel B: Continuous Data}} & &\\
  500 & 5 & 2 & 2 & continuous & 0.046 sec & 0.005 sec & 0.028 sec & 0.007 sec \\ 
  1000 & 5 & 2 & 2 & continuous & 0.176 sec & 0.019 sec & 0.089 sec & 0.022 sec \\ 
  2000 & 5 & 2 & 2 & continuous & 0.691 sec & 0.076 sec & 0.319 sec & 0.071 sec \\ 
    \addlinespace
  500 & 10 & 2 & 2 & continuous & 0.179 sec & 0.02 sec & 0.082 sec & 0.022 sec \\ 
  1000 & 10 & 2 & 2 & continuous & 0.698 sec & 0.077 sec & 0.287 sec & 0.07 sec \\ 
  2000 & 10 & 2 & 2 & continuous & 2.846 sec & 0.316 sec & 1.019 sec & 0.225 sec \\ 
    \addlinespace
  500 & 5 & 3 & 2 & continuous & 0.06 sec & 0.007 sec & 0.024 sec & 0.004 sec \\ 
  1000 & 5 & 3 & 2 & continuous & 0.227 sec & 0.025 sec & 0.068 sec & 0.013 sec \\ 
  2000 & 5 & 3 & 2 & continuous & 0.897 sec & 0.098 sec & 0.221 sec & 0.065 sec \\ 
    \addlinespace
  500 & 10 & 3 & 2 & continuous & 0.229 sec & 0.025 sec & 0.061 sec & 0.01 sec \\ 
  1000 & 10 & 3 & 2 & continuous & 0.9 sec & 0.099 sec & 0.188 sec & 0.03 sec \\ 
  2000 & 10 & 3 & 2 & continuous & 3.611 sec & 0.404 sec & 0.572 sec & 0.107 sec \\ 
    \addlinespace
  500 & 5 & 2 & 3 & continuous & 1.266 min & 7.673 sec & 11.209 sec & 2.877 sec \\ 
  1000 & 5 & 2 & 3 & continuous & 9.702 min & 50.413 sec & 1.205 min & 20.809 sec \\ 
  2000 & 5 & 2 & 3 & continuous & 1.227 hrs & 5.838 min & 8.053 min & 2.113 min \\ 
    \addlinespace
  500 & 10 & 2 & 3 & continuous & 9.413 min & 46.492 sec & 1.141 min & 21.284 sec \\ 
  1000 & 10 & 2 & 3 & continuous & 1.185 hrs & 3.211 min & 7.562 min & 2.432 min \\ 
  2000 & 10 & 2 & 3 & continuous & 9.273 hrs & 29.096 min & 46.807 min & 13.962 min \\ 
    \addlinespace
  500 & 5 & 3 & 3 & continuous & 1.443 min & 4.879 sec & 7.722 sec & 1.245 sec \\ 
  1000 & 5 & 3 & 3 & continuous & 11.239 min & 37.42 sec & 37.187 sec & 6.663 sec \\ 
  2000 & 5 & 3 & 3 & continuous & 1.478 hrs & 4.908 min & 3.224 min & 46.965 sec \\ 
    \addlinespace
  500 & 10 & 3 & 3 & continuous & 11.502 min & 40.822 sec & 35.146 sec & 6.565 sec \\ 
  1000 & 10 & 3 & 3 & continuous & 1.516 hrs & 5.657 min & 2.885 min & 34.976 sec \\ 
  2000 & 10 & 3 & 3 & continuous & 12.06 hrs & 31.1 min & 14.442 min & 3.183 min \\ 
  \label{tab:ptdefault}
\end{longtable}

\doublespacing

In all cases {\tt fastpolicytree} performs orders of magnitudes more quickly, with the difference becoming more pronounced for deeper trees. The difference in performance is particularly large when using discrete covariates. For example for the $(N=10000,p=60,\mathrm{Acts}=2,\mathrm{Depth}=3)$ configuration on discrete covariates, \pt{} takes 40.86 minutes (on average) whereas \fpt{} takes 5.452 seconds. This is a 449\% speed-up. 

The \fpt{} performance improvement on continuous data is less dramatic, but still good. For example, a depth-three tree learned from 10 continuous covariates where N=2000 with 3 treatment actions takes over 12 hours using the {\tt policytree} package. With {\tt fastpolicytree}, the time reduces to 14.4 minutes, roughly 50 times faster.\footnote{The learned policy $\pi$ and the estimated policy values $\mathbb{E}[\Gamma_{iw}(\pi(X_i))]$ are identical for both versions and are not reported here.}  

\subsection{Using Split Steps}

We now turn to a comparison between our {\tt fastpolicytree} and a proposed technique to reduce the runtime of the original {\tt policytree} package. Increasing the optional approximation parameter, {\tt split.step}, decreases the number of possible splits to consider during the tree search. The authors of {\tt policytree} demonstrate that increasing this parameter for learning trees with continuous covariates can greatly improve the runtime compared to the default {\tt split.step} value of 1. However, less is known about the potential loss in accuracy of the estimated policy rewards due to bias in the learned trees occurring when samples are skipped (i.e., trees are no longer guaranteed \textit{optimal}). As in the first simulation study, we compare the runtimes of the two packages, however for the {\tt policytree} package we set the {\tt split.step} parameter equal to 10. We report the runtimes and also the root mean squared error (RMSE) between the {\tt fastpolicytree} and {\tt policytree} estimated policy values, i.e.:

\begin{equation*}
    RMSE = \sqrt{\frac{1}{nsim}\sum(\mathbb{E}[\Gamma_{iw}(\pi_{FPT}(X_i))] - \mathbb{E}[\Gamma_{iw}(\pi_{PT}(X_i))])^2}
\end{equation*}

\singlespacing

\begin{table}[t]
\centering
\caption{Simulation results: splitstep policytree (Continuous covariate data)}
\begin{tabular}{lccccccccc} 
\toprule
\multicolumn{4}{c}{\textbf{ }} & \multicolumn{2}{c}{\textbf{PT}} & \multicolumn{2}{c}{\textbf{FPT}} & \textbf{RMSE}\\
\cmidrule{5-6} \cmidrule{7-8}
N & p & Acts & Depth & Time & SD & Time & SD \\
\midrule
1000 & 30 & 3 & 2 & 0.836 sec & 0.081 sec & 1.287 sec & 0.21 sec & 0.003 \\ 
  10000 & 30 & 3 & 2 & 1.6 min & 30.199 sec & 1.38 min & 27.151 sec & 0.000 \\ 
  100000 & 30 & 3 & 2 & 8.363 hrs & 2.935 hrs & 2.32 hrs & 41.997 min & 0.000 \\ 
  1000 & 30 & 20 & 2 & 2.423 sec & 0.156 sec & 3.988 sec & 0.506 sec & 0.005 \\ 
  10000 & 30 & 20 & 2 & 5.94 min & 1.085 min & 4.879 min & 38.474 sec & 0.000 \\ 
  100000 & 30 & 20 & 2 & 43.51 hrs & 9.187 hrs & 11.573 hrs & 3.013 hrs & 0.000 \\ 
\bottomrule
\end{tabular} 
\caption*{\textit{Note: } Results are averaged across 100 repetitions. The splitting step is set to 10 for all Athey {\tt policytree} versions. Reward RMSE is calculted as the square root of the mean squared difference between FPT and {\tt policytree} rewards.} 

\end{table}

\doublespacing

\begin{figure}
    \centering
    \caption{splitstep: Distribution of Errors}
    \includegraphics[width=\linewidth]{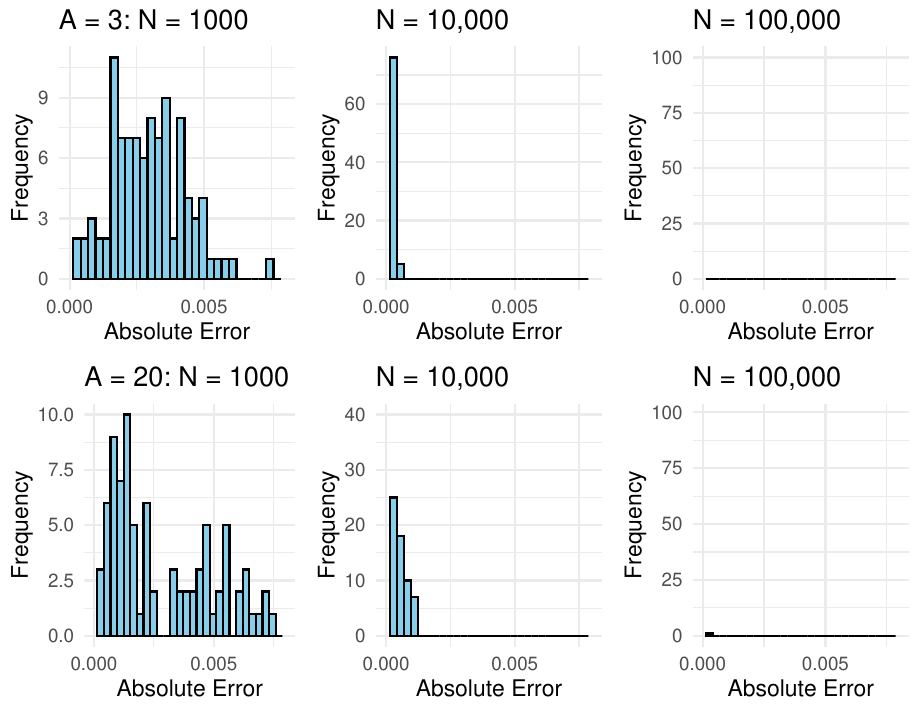}
    \label{fig:splitstep_errors}
    \caption*{\textit{Note: } Results are across 100 repetitions of the continuous simulations. The splitting step is set to 10 for all Athey {\tt policytree} versions. Absolute error is calculated as difference between FPT and {\tt policytree} rewards.}
\end{figure}

Even with the {\tt split.step} parameter increased, the {\tt fastpolicytree} runtimes are an improvement over the original package version in larger samples. For $N=100,000$ with 30 continuous covariates and 20 actions, a depth two tree learned using {\tt fastpolicytree} is roughly four times faster. For smaller samples the runtimes are similar, but there is a loss in accuracy between the two versions, indicating that using the {\tt fastpolicytree} may be preferable when accuracy in the predicted reward is desirable. We depict this graphically in Figure \ref{fig:splitstep_errors}, which shows the distribution of absolute errors between the {\tt fastpolicytree} and the {\tt policytree} with {\tt split.step} parameter increased. 

\section{Conclusion and Future Work}

In this paper, we describe an algorithm, \fpt, for learning optimal policy trees that exploits a number of optimisations to improve run times significantly when compared to \pt. We have performed a large number of benchmarking experiments to support the claim of improved running time. With this faster algorithm it is possible to learn optimal policy trees of greater depth, thus helping practitioners find better policies by accounting for more individual characteristics in the treatment allocation procedure. 

The methods employed by \fpt{} (the use of bounds, caching, etc) are standard ones in discrete optimisation although they have been specialised for the particular problem (learning optimal policy trees) at hand. As well as algorithmic improvements, our implementation work also makes a difference according to covariate type. Discrete covariates can take advantage of our `Method~2' set implementation (see Section~\ref{sec:altset}) which explains the particular good runtimes we have for them.

Discrete optimisation is a well-developed and active area of research and we expect further effort in applying its methods to policy tree learning to bear fruit. Indeed, there is considerable work specifically on learning `optimal trees'. In many cases the optimal trees being learned are classification or regression trees rather than policy trees but there is considerable commonality between these tasks. \citet{NEURIPS2023_1d5fce96} provide a good summary of recent work in this area and describe STreeD (Separable
Trees with Dynamic programming), a method which has good results for both (cost-sensitive) classification trees and policy trees.



The \fpt{} R package is available at : \verb+https://github.com/jcussens/tailoring+. Three types of the package are available: a source package, a Windows binary package and a MacOS binary package. The code is written in C with some `wrapper' code written in R. It is also possible to create a standalone 
Linux executable (if one has a C compiler installed). The R package contains a single function: {\tt fastpolicytree}, which is intended as a direct replacement for the {\tt policy\_tree} function provided by the \pt{} R package. The {\tt fastpolicytree} function has additional arguments which can be used to alter its standard approach to finding optimal policy trees, but a user can safely leave these at their default values. The {\tt fastpolicytree} function lacks the {\tt split.step} argument that the {\tt policy\_tree} function has, since we focus on finding only optimal trees (for a given depth). The \pt{} R package has many useful functions in addition to {\tt policy\_tree}. We have not replicated these in the \fpt{} package, so in practice it makes sense to have both packages available.
\clearpage 
\bibliography{Bibliography/paperpile,Bibliography/extra}
\end{document}